\documentclass[aps,prd,preprint,superscriptaddress,tightenlines,nofootinbib]{revtex4}

\usepackage{graphicx}
\usepackage{dcolumn}
\usepackage{bm}

\begin{document}


\begin{flushright}
CLNS 04-1887\\
CLEO 04-11\\
CLEO CONF 04-04\\
ICHEP04 ABS10-0774\\
\end{flushright}

\title{Measurement of the Muonic Branching Fraction of the\\ 
Narrow $\Upsilon$ Resonances at CLEO}
\thanks{Submitted to the 32$^{\rm nd}$ International Conference on High Energy
Physics, Aug 2004, Beijing}


\author{G.~S.~Adams}
\author{M.~Chasse}
\author{M.~Cravey}
\author{J.~P.~Cummings}
\author{I.~Danko}
\author{J.~Napolitano}
\affiliation{Rensselaer Polytechnic Institute, Troy, New York 12180}
\author{D.~Cronin-Hennessy}
\author{C.~S.~Park}
\author{W.~Park}
\author{J.~B.~Thayer}
\author{E.~H.~Thorndike}
\affiliation{University of Rochester, Rochester, New York 14627}
% ...
\collaboration{CLEO Collaboration} 
\noaffiliation


\date{August 1, 2004}

\begin{abstract}
The decay branching fractions of the three narrow $\Upsilon$ resonances to $\mu^+ \mu^-$ have been measured by analyzing about $4.3$ fb$^{-1}$ $e^+e^-$ data collected in the vicinity of the resonances with the CLEO III detector. The branching fraction ${\cal B}(\Upsilon(1S) \to \mu^+ \mu^-) = (2.49 \pm 0.02 \pm 0.07)\%$ is consistent with the current world average but ${\cal B}(\Upsilon(2S) \to \mu^+ \mu^-) = (2.03 \pm 0.03 \pm 0.08)\%$ and ${\cal B}(\Upsilon(3S) \to \mu^+ \mu^-) = (2.39 \pm 0.07 \pm 0.10)\%$ are significantly larger than prior results. These new muonic branching fractions imply a narrower total decay width for the $\Upsilon$(2S) and $\Upsilon$(3S) resonances and lower other branching fractions that rely on these decays in their determination.
\end{abstract}

\maketitle

\section{Introduction}

The main theoretical obstacle in determining the amount of CP-violation that comes from the Standard Model is related to uncertainties in computing various hadronic quantities. 
Recent advances in Lattice QCD promise very accurate predictions for a wide variety of non-perturbative processes \cite{preciseLQCD}. 
However, substantially improved data are needed to confront these predictions.
The long-lived $b\bar{b}$ states are especially well suited for establishing the accuracy of Lattice QCD calculations \cite{LQCD} as well as testing effective theories of the strong interactions, such as potential models \cite{Yreview}, in the heavy quark sector.
The large data sample of about $4.5$ fb$^{-1}$ collected recently by the CLEO detector in the vicinity of the $\Upsilon$(nS) ($n=1, 2, 3$) resonances enables us to determine the $b\bar{b}$ resonance parameters, such as the leptonic and total decay widths, with unprecedented precision. 

The total widths of the narrow $\Upsilon$ resonances below the open-bottom threshold cannot be measured directly since their natural widths ($25-50$ keV) are much narrower than the collider beam energy spread ($4-5$ MeV). 
The common indirect method to determine the total decay width ($\Gamma$) is to combine the leptonic branching fraction (${\cal B}_{\ell\ell}$) with the leptonic decay width ($\Gamma_{\ell\ell}$): $\Gamma = \Gamma_{\ell\ell}/{\cal B}_{\ell\ell}$ \cite{Yreview,PDG}.
In practice, assuming lepton universality, the leptonic decay width is replaced by $\Gamma_{ee}$, which can be extracted from the energy-integrated resonant hadron production cross section in $e^+e^-$ collisions, while the leptonic branching fraction is replaced by the muonic branching fraction, ${\cal B}_{\mu\mu} \equiv {\cal B}(\Upsilon \to \mu^+\mu^-$), which can be measured more accurately than ${\cal B}_{ee}$ or ${\cal B}_{\tau\tau}$. 
Therefore, it is very important to measure ${\cal B}_{\mu\mu}$ precisely in order to determine the total decay widths of the narrow $\Upsilon$ resonances.

The leptonic branching fraction is also interesting in its own right since it represents the strength of the $\Upsilon$ decay to lepton pairs via annihilation to a virtual photon relative to the decay into $ggg$ and $gg\gamma$.
Furthermore, ${\cal B}_{\mu\mu}$ is generally used in determinations of the branching fractions of hadronic and electromagnetic transitions among the $\Upsilon$ states since these decays are often measured by observing the decay of the lower lying resonances to lepton pairs.
In addition, comparing ${\cal B}_{\mu\mu}$ to ${\cal B}_{ee}$ as well as ${\cal B}_{\tau\tau}$ can provide a check of lepton universality and test the possible existence of new physics beyond the Standard Model \cite{Higgs}. 

Based on previous measurements, ${\cal B}_{\mu\mu}$ is established with a 2.4\% accuracy for the $\Upsilon$(1S) \cite{PDG}, and a modest 16\% and 9\% accuracy for the $\Upsilon$(2S) \cite{CLEO_2S,ARGUS,CUSB,CBAL} and $\Upsilon$(3S) \cite{CLEO_3Sa,CUSB,CLEO_3Sb}, respectively.
This paper reports the measurement of ${\cal B}_{\mu\mu}$ for all three narrow resonances with a much larger data set and a more advanced detector. 
The new results enable us to determine the total decay widths of the $\Upsilon$(2S) and $\Upsilon$(3S) with better precision.

\section{Data sample and analysis technique}

The data used in this analysis were collected by the CLEO III detector during 2001-2002 at the Cornell Electron Storage Ring, a symmetric $e^+ e^-$ collider.
The analysis relies on the excellent charged particle tracking, electromagnetic calorimetry, and muon identification of CLEO III.
The upgraded tracking system consists of a new 4-layer double-sided Silicon vertex detector and a new 47-layer drift chamber \cite{DR3} residing in a 1.5 T solenoidal magnetic field.
The CsI(Tl) crystal calorimeter and the muon detector system inherited from CLEO II \cite{CLEOII} can identify muons with momentum above 1.0 GeV/$c$ with high efficiency.

To determine ${\cal B}_{\mu\mu}$, we measure the $\Upsilon \to \mu^+\mu^-$ decay rate relative to the $\Upsilon \to hadron$ rate 
\begin{equation}
\tilde{\cal{B}} \equiv \frac{\Gamma_{\mu\mu}}{\Gamma_{\rm had}} 
= \frac{N(\Upsilon \to \mu^+\mu^-)}{N(\Upsilon \to hadron)}
=\frac{\tilde{N}_{\mu\mu}/\varepsilon_{\mu\mu}}{\tilde{N}_{\rm had}/\varepsilon_{\rm had}},
\end{equation}
where $\Gamma_{\rm had}$ includes all decay modes of the resonances other than $e^+e^-$, $\mu^+\mu^-$, and $\tau^+\tau^-$.
Assuming lepton universality, {\it i.e.} $\Gamma_{ee} = \Gamma_{\mu\mu} = \Gamma_{\tau\tau}$, we have
\begin{equation}
{\cal B}_{\mu\mu} \equiv \frac{\Gamma_{\mu\mu}}{\Gamma}
= \frac{\Gamma_{\mu\mu}}{\Gamma_{\rm had}(1+3{\tilde{\cal{B}}}_{\mu\mu})}
= \frac{{\tilde{\cal{B}}}_{\mu\mu}}{(1+3{\tilde{\cal{B}}}_{\mu\mu})}.
\end{equation}

The major source of background is non-resonant (continuum) production of $\mu^+\mu^-$ and hadrons via $e^+ e^- \to \mu^+\mu^-$ and $e^+ e^- \to q\bar{q}$ ($q=u,d,c,s$), respectively, which cannot be distinguished experimentally from the corresponding resonance decays.
Hence, we use continuum data collected at energies just below each resonance  to subtract these backgrounds. 
Then the number of $\Upsilon$ decays to $\mu^+\mu^-$ (or hadron) is $\tilde{N} = \tilde{N}_{on} - S \tilde{N}_{off}$, where $S$ scales the luminosity of the off-resonance data to that of the on-resonance data and accounts for the $1/s$ dependence of the cross section.

Background contributions from non-resonant $e^+e^- \to \tau^+\tau^-$, two-photon fusion ($e^+e^- \to e^+e^- \gamma^{\star}\gamma^{\star}$), or from radiative return to the lower resonances are less than 0.2\% after the off-resonance subtraction.
The remaining backgrounds (to $\mu^+\mu^-$) are mainly from cosmic rays, and more importantly from $\Upsilon$(3S) and $\Upsilon$(2S) decays to a lower $\Upsilon$ state, in which the lower resonance decays to $\mu^+\mu^-$ and the extra particles escape detection. 
The decay $\Upsilon \to \tau^+\tau^-$ can also contribute to both the $\mu^+\mu^-$ and hadron measurements. 

Our results are based upon 1.1 fb$^{-1}$ (1S), 1.2 fb$^{-1}$ (2S), and 1.2 fb$^{-1}$ (3S) data collected within 2-3 MeV at the peak of each resonance (``on-resonance samples'') as well as off-resonance samples which were collected 20-30 MeV below the resonances and represent 0.19 fb$^{-1}$ (below 1S), 0.44 fb$^{-1}$ (below 2S), and 0.16 fb$^{-1}$ (below 3S).
The scale factors, $S$, between the on-resonance and the corresponding off-resonance samples are calculated from the luminosity measured with $e^+ e^- \to \gamma\gamma$ process \cite{lumi} which, unlike the $e^+ e^-$ final state, is not contaminated by resonance decays.

\section{Selection of $\Upsilon \to \mu^+\mu^-$ events}

We select $\mu^+\mu^-$ events by requiring exactly two oppositely charged tracks, each with momentum between 70\% and 115\% of the beam energy ($E_{\rm beam}$), with polar angle $| \cos\theta | < 0.8$, and with opening angle of the tracks grater than 170$^{\circ}$.
Muon identification requires each track to deposit $0.1-0.6$ GeV in the electromagnetic calorimeter, characteristic of a minimum ionizing particle, and at least one track to penetrate deeper than five interaction lengths into the muon chambers.

\begin{figure}
\includegraphics*[width=5.0in]{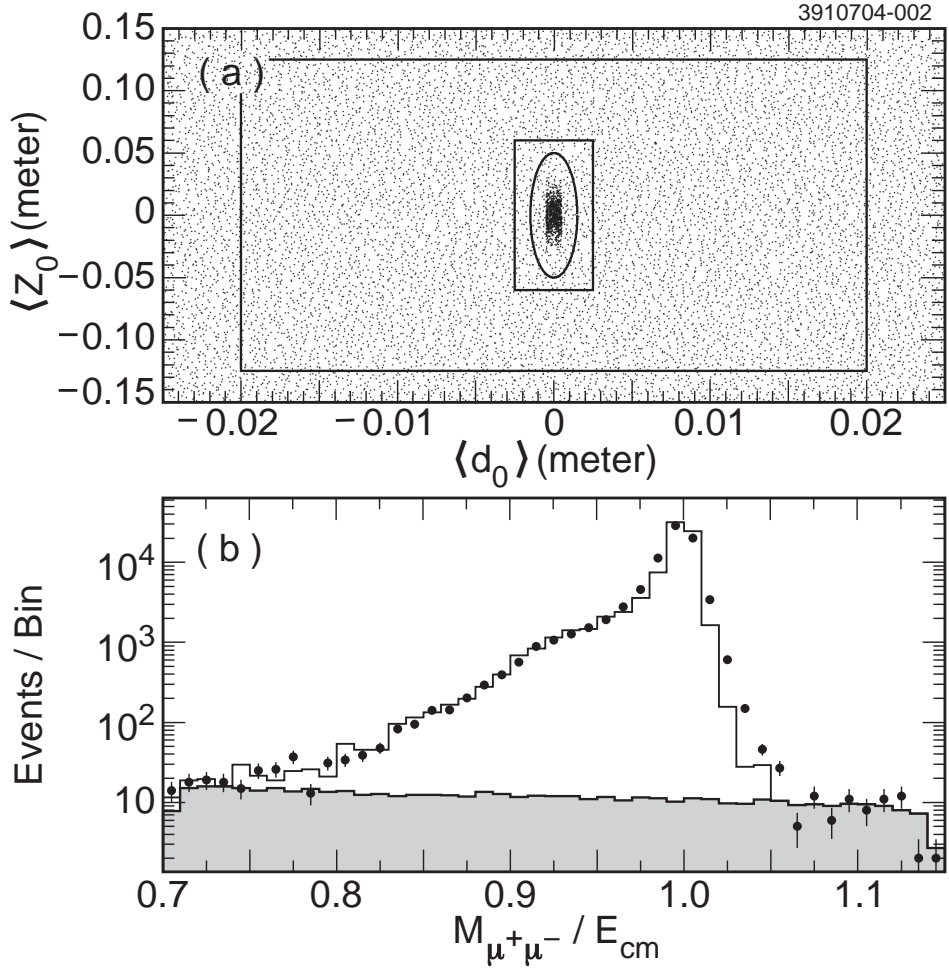}
\caption{a) Distribution of $\mu^+\mu^-$ candidate events in off-resonance data below the $\Upsilon$(3S) over the $\langle z_0 \rangle$ vs. $\langle d_0 \rangle$ plane. The ellipse encircles the signal region while the two rectangles enclose the sideband. b) Invariant mass distribution of the $\mu^+\mu^-$ candidates in the signal region (dots) overlaid with the expected distribution from Monte Carlo simulation of $e^+e^- \to \mu^+\mu^-$ events. The shaded histogram represents the scaled distribution for events in the sideband. The vertical scale is logarithmic.}
\label{fig:cosmic}
\end{figure}

We control the cosmic-ray background using the track impact parameters with respect to the $e^+e^-$ interaction point (beam spot).
From the location of the point nearest to the beam spot (as seen in the $r-\phi$ plane) on both tracks we calculate the separation between the two tracks along the beam axis ($\Delta z_0$) and in the perpendicular plane ($\Delta d_0$) as well as their average distance from the beam spot along the beam axis ($\langle z_0 \rangle$) and in the perpendicular plane ($\langle d_0 \rangle$).
We require $| \Delta z_0 | < 5$cm, $| \Delta d_0 | < 2$mm and
$(\langle z_0 \rangle/5cm)^2 + (\langle d_0 \rangle/1.5mm)^2 <1$.
Cosmic events are uniformly distributed in the $\langle z_0 \rangle$ and $\langle d_0 \rangle$ variables while the events from $e^+e^-$ collision populate a small area around $(\langle z_0 \rangle, \langle d_0 \rangle) = (0,0)$. 
We use a two-dimensional sideband around the signal region to estimate the remaining cosmic-ray background (Fig.~\ref{fig:cosmic}a). 
This background is $0.3-0.6$\% depending on the data sample.
The observed rate of events with $M_{\mu\mu}>1.1E_{\rm cm}$ (after the momentum cuts have been relaxed) is consistent within 10\% with this background estimate (Fig.~\ref{fig:cosmic}b).
We correct the number of $\mu^+\mu^-$ events observed in the on-resonance and off-resonance samples individually for the cosmic background.

\begin{figure}
\includegraphics*[width=5.0in]{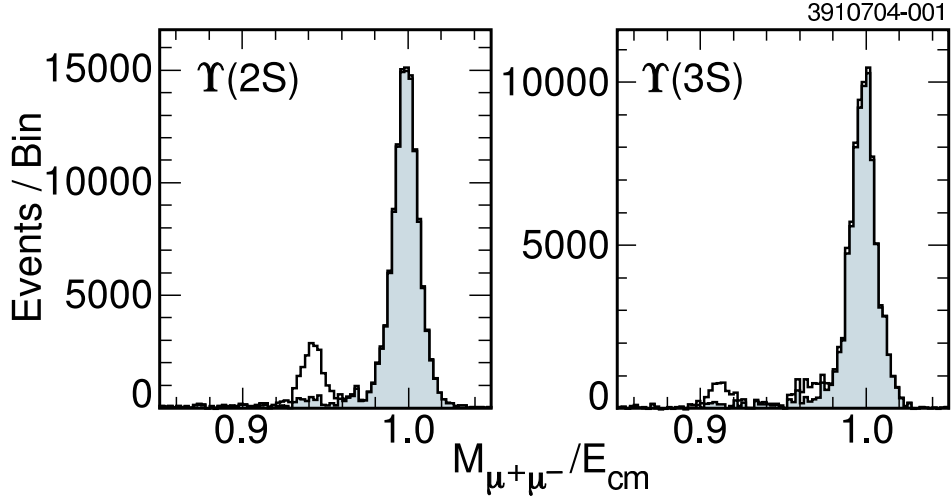}
\caption{Distribution of the invariant mass of $\mu^+\mu^-$ candidates from the $\Upsilon$(2S) (left) and $\Upsilon$(3S) (right) after off-resonance subtraction. The empty (shaded) histograms show the distributions before (after) rejecting events with extra showers in the calorimeter.}
\label{fig:cascade}
\end{figure}

Requiring exactly two charged tracks suppresses the indirect $\mu^+\mu^-$ production at the $\Upsilon$(2S) and $\Upsilon$(3S) from $\Upsilon(nS) \to \Upsilon(mS)\pi^+\pi^-$ followed by $\Upsilon(mS) \to \mu^+\mu^-$ but it is ineffective against cascade decays containing only neutral particles.
To reduce this background, we require fewer than two extra showers with more than 50 MeV (100 MeV) energy in the barrel (endcap) section of the calorimeter.
This requirement significantly suppresses the background while the direct muon efficiency changes by less than 1\% (Fig.~\ref{fig:cascade}).
We estimate the remaining cascade background using measured branching fractions \cite{PDG} and a Monte Carlo simulation.
The residual cascade background is ($2.9\pm 1.5$)\% and ($2.2\pm 0.7$)\% for $\Upsilon$(2S) and $\Upsilon$(3S), respectively, where the uncertainty is dominated by the decay branching fractions and the selection efficiencies.

\begin{figure}
\includegraphics*[width=5.0in]{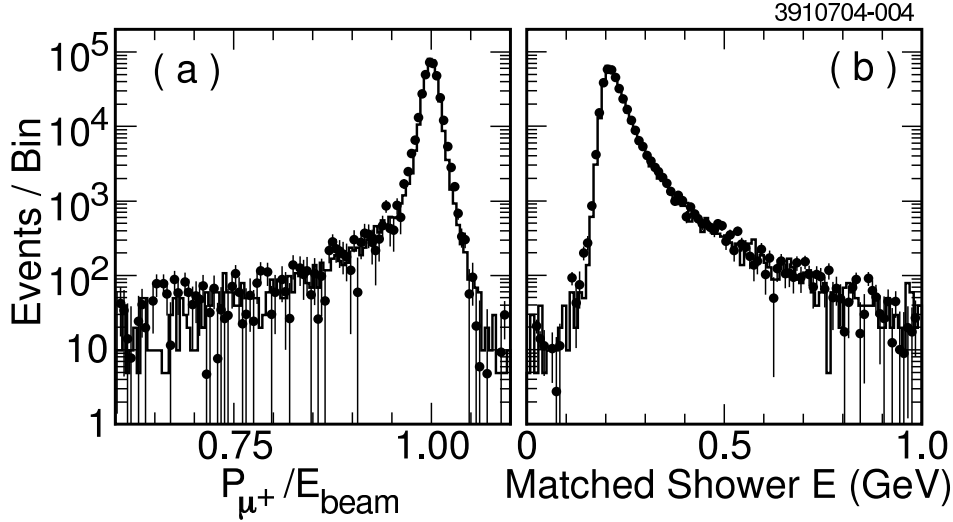}
\caption{Distribution of the momentum (left) and the shower energy (right) of the $\mu^+$ candidate from $\Upsilon$(1S) resonance decay after off-resonance subtraction (dots) and in resonance Monte Carlo simulation (histogram). The vertical scale is logarithmic.}
\label{fig:data_mc}
\end{figure}

The overall selection efficiency for $\Upsilon \to \mu^+\mu^-$ decays is ($65.2 \pm 1.2$)\% from a GEANT-based Monte Carlo simulation.
The relative systematic uncertainty on the efficiency is 1.8\% which is dominated by the uncertainty in the detector simulation (1.7\%) determined from a detailed comparison between data and Monte Carlo distributions of all selection variables (Fig.~\ref{fig:data_mc}).
The $\mu^+\mu^-$ event selection has been checked by calculating the $e^+e^- \to \mu^+\mu^-$ cross section using the number of $\mu^+\mu^-$ events observed in the off-resonance samples, the corresponding Monte Carlo efficiency, and the Bhabha luminosity.
The measured cross section is consistent with the theoretical cross section including higher order radiative corrections \cite{FPAIR}.

\section{Selection of $\Upsilon \to hadron$ events}

When selecting hadronic events, we minimize the systematic uncertainty by maintaining high selection efficiency.
QED backgrounds ($e^+e^- \to e^+e^-/\mu^+\mu^-/\gamma\gamma$) are suppressed by requiring $\geq3$ charged particles. 
In addition, for low multiplicity events with $<5$ charged tracks, we require the total energy detected in the electromagnetic calorimeter to be more than 15\% of the center of mass energy ($E_{\rm cm}$), and either the total calorimeter energy to be less than 75\% of $E_{\rm cm}$ or the most energetic shower to be less than 75\% of $E_{\rm beam}$.
To suppress beam-gas and beam-wall interactions we reject events in which the total energy visible in the calorimeter and in the tracking system is less than 20\% of $E_{\rm cm}$. 
We also use the event vertex position to reject the beam-related background as well as cosmic rays and to estimate the residual background from these sources.
Background to the hadrons from $\Upsilon \to \tau^+\tau^-$ decay is estimated to be ($0.7 \pm 0.1$)\%, ($0.4 \pm 0.3$)\%, and ($0.5\pm 0.2$)\% for the $\Upsilon$(1S), $\Upsilon$(2S), and $\Upsilon$(3S), respectively, where the uncertainty is dominated by the inaccuracy in the leptonic branching fractions and in theMonte Carlo detection efficiency.

We determine the hadron selection efficiency for $\Upsilon \to hadron$ decays from Monte Carlo simulation of the detector using event generators based on Jetset 7.3 and 7.4 \cite{Jetset74}.
The two simulations result in a slightly different selection efficiency and a comparison of the distributions of the selection variables in data and Monte Carlo suggests that the real acceptance is apparently in between. 
Hence, we average the two Monte Carlo efficiencies and assign a relative systematic uncertainty to cover the difference between the two simulations: 1.6\% (1S), 1.2\% (2S) and 1.3\% (3S).
Uncertainties in the cascade and leptonic branching fractions of the $\Upsilon$ resonances contribute with an additional 0.3\% added in quadrature to the total efficiency uncertainty in case of the $\Upsilon$(2S) and $\Upsilon$(3S).
The overall acceptance for hadronic resonance decays varies between 96-98\% (largest for the 1S and smallest for the 2S) depending on the relative rate of cascade decays that can produce a stiff $e^+e^-$ or $\mu^+\mu^-$ in the final state (see Table~\ref{tab:result}).

\section{Branching fractions and total decay widths}

\begin{figure}
\includegraphics*[width=5.0in]{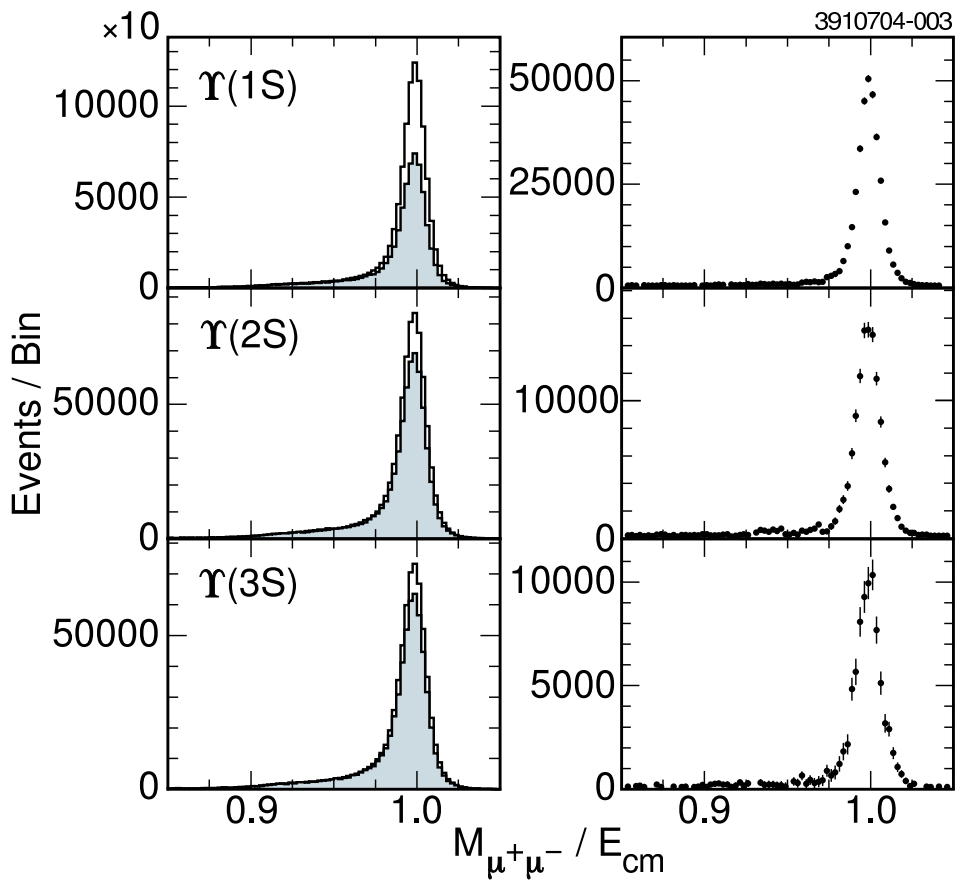}
\caption{Mu-pair invariant mass in on-resonance (empty) and scaled off-resonance (shaded) data on the left and in off-resonance subtracted resonance data on the right. }
\label{fig:muons}
\end{figure}

Table~\ref{tab:result} presents the observed number of $\mu^+\mu^-$ and hadronic events from resonance decays ($\tilde{N}$), the corresponding selection efficiencies ($\varepsilon$) along with the statistical uncertainty. 
The invariant mass distribution of the $\mu^+\mu^-$ candidates in the on-resonance and off-resonance samples and after off-resonance subtraction are shown in Fig.~\ref{fig:muons}.

\begin{table}[h]
  \centering
  \caption{Number of resonance decays to $\mu^+\mu^-$ and hadrons ($\tilde{N}$), selection efficiencies ($\varepsilon$), and the muonic branching fractions ($\cal{B}$). $\cal{B}_{\mu\mu}$ is corrected for interference. The error is statistical only.}\label{tab:result}
\bigskip
\begin{tabular}[c]{lccc}
  \hline
  \hline
    & $\Upsilon$(1S) & $\Upsilon$(2S) & $\Upsilon$(3S) \\
  \hline
   $\tilde{N}_{\mu\mu}$ & $344908\pm2485$ & $119588\pm1837$ & $81179\pm2660$ \\
   $\varepsilon_{\mu\mu}^{r}$ & $0.652\pm0.002$ & $0.652\pm0.002$ & $0.652\pm0.002$ \\
   $\tilde{N}_{\rm had}$ & $18957575\pm11729$ & $7838270\pm8803$ & $4641369\pm12645$ \\
   $\varepsilon_{\rm had}^{r}$ & $0.979$ & $0.965$ & $0.975$ \\
  \hline
   ${\cal B}_{\mu\mu}$ & $0.0249\pm0.0002$ & $0.0203\pm0.0003$ & $0.0239\pm0.0007$ \\
  \hline
  \hline
\end{tabular}
\end{table}

Since the effect of the interference between the resonant and non-resonant production is energy dependent (expected to be destructive (constructive) below (above) the resonance \cite{interference}) and its relative contribution to $\mu^+\mu^-$ is about ten times larger than to hadrons, the measured branching fraction depends slightly on the center of mass energy at which the data were taken.
We estimate the effect of interference at the luminosity-weighted average center of mass energies of the on-resonance as well as the off-resonance samples using the convolution of an interference-corrected Breit-Wigner resonance shape, a Gaussian energy spread, and a radiative tail \cite{Kuraev-Fadin}.
The resulting correction factors to the observed $\cal{B}_{\mu\mu}$ branching fractions due to interference are 0.984, 0.961, and 0.982 for the $\Upsilon$(1S), $\Upsilon$(2S), and $\Upsilon$(3S), respectively.
The $\Upsilon(nS) \to \mu^+\mu^-$ branching fractions listed in Table~\ref{tab:result} are corrected for the interference.

The total fractional systematic uncertainties in ${\cal B}_{\mu\mu}$ are 2.7\% (1S), 3.7\% (2S), and 4.1\% (3S), respectively.
They are the quadrature sums of the fractional uncertainties due to several sources listed in Table \ref{tab:systematics}.
The systematic uncertainty due to detection efficiency ($\varepsilon$) is from detector modeling (dominant), trigger efficiency and Monte Carlo statistics.
The systematic uncertainty in the raw event number ($\tilde{N}$) is due to the uncertainties in various backgrounds.
Uncertainties in the interference calculation and variations in the center of mass energy contribute 1\%.
The dominant source of the systematic uncertainty in the cases of $\Upsilon$(2S) and $\Upsilon$(3S) is due to the uncertainty in the scale factor between the on-resonance and off-resonance data.

\begin{table}[h]
  \centering
  \caption{Fractional systematic uncertainties (\%) in ${\cal B}_{\mu\mu}$.}
  \label{tab:systematics}
\bigskip
\begin{tabular}[c]{lccc}
  \hline
  \hline
    & $\Upsilon$(1S) & $\Upsilon$(2S) & $\Upsilon$(3S) \\
  \hline
   $\varepsilon_{\rm had}^{r}$ & 1.6 & 1.3 & 1.4 \\
   $\tilde{N}_{\rm had}(res)$ & 0.2 & 0.3 & 0.4 \\
   $\varepsilon_{\mu\mu}^{r}$ & 1.8 & 1.8 & 1.8 \\
   $\tilde{N}_{\mu\mu}(res)$ & 0.1 & 1.6 & 0.9 \\
   Scale factor & 0.8 & 2.3 & 3.1 \\
   Interference & 1 & 1 & 1 \\
  \hline
   Total error & 2.7 & 3.7 & 4.1 \\
 \hline
 \hline
\end{tabular}
\end{table}

\begin{figure}
  \centering
  \includegraphics*[width=6.0in]{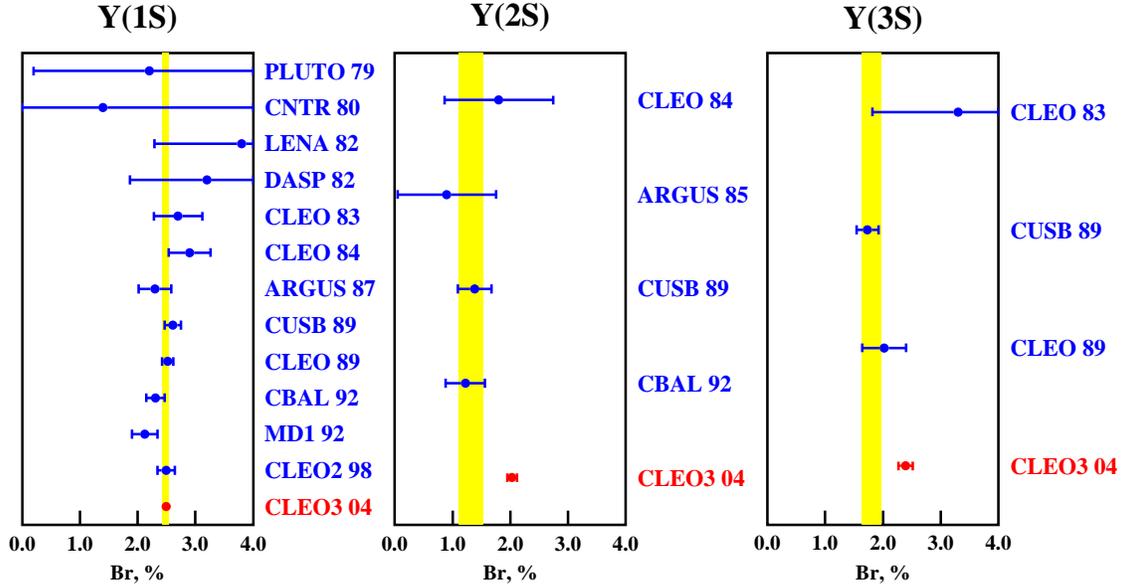}
  \caption{Comparison of the muonic branching fractions obtained from this analysis with earlier measurements \cite{PDG} and with the PDG averages represented by the vertical bands. The error bars represent the statistical and systematic errors combined in quadrature.}\label{fig:Bmumu_comp}
\end{figure}

The final muonic branching fractions, including systematic uncertainties, are
\begin{equation}
\begin{array}{c}
{\cal B}_{\mu\mu}(1S) = (2.49 \pm 0.02 \pm 0.07) \%,\\
{\cal B}_{\mu\mu}(2S) = (2.03 \pm 0.03 \pm 0.08) \%,\\
{\cal B}_{\mu\mu}(3S) = (2.39 \pm 0.07 \pm 0.10) \%.\\
\end{array}
\end{equation}
The result for the $\Upsilon$(1S) is in very good agreement with the current world average \cite{PDG}, while our $\Upsilon$(2S) and $\Upsilon$(3S) results are significantly larger than earlier results of the CUSB \cite{CUSB} and Crystal Ball \cite{CBAL} experiments (Fig.~\ref{fig:Bmumu_comp}).

The total decay widths of the resonances can be expressed as \cite{PDG}
\begin{equation}
\Gamma = \frac{\Gamma_{ee}\Gamma_{\rm had}/\Gamma}{{\cal B}_{\mu\mu}(1 - 3{\cal B}_{\mu\mu})}
\end{equation}
Our improved muonic branching fractions combined with the current values of $\Gamma_{ee}\Gamma_{\rm had}/\Gamma$ \cite{PDG} lead to the following new values for the total decay widths of the three narrow $\Upsilon$ resonances:
\begin{equation}
\begin{array}{c}
 \Gamma (1S) = (52.8 \pm 1.8) {\rm keV},\\
 \Gamma (2S) = (29.0 \pm 1.6) {\rm keV},\\
 \Gamma (3S) = (20.3 \pm 2.1) {\rm keV}.\\
\end{array}
\end{equation}

The new total widths of the $\Upsilon$(2S) and $\Upsilon$(3S) have a significant impact on the comparison between theoretical and experimental values of hadronic and radiative widths of these resonances since the experimental widths are determined as a product of the total widths and the measured transition branching fractions.
The new value of ${\cal B}_{\mu\mu}$(2S) also significantly alters ${\cal B}(\Upsilon(3S) \to \pi\pi\Upsilon(2S))$ and ${\cal B}(\Upsilon(3S) \to \gamma\gamma\Upsilon(2S))$ (consequently ${\cal B}(\chi_b(2P_J) \to \gamma\Upsilon(2S))$ as well) which are extracted from the measured exclusive $\Upsilon(3S) \to \pi\pi\ell^+\ell^-$ and $\gamma\gamma \ell^+\ell^-$ branching fractions.

\section{Conclusion}

We have measured the muonic branching fraction of the narrow $\Upsilon$ resonances below the open-bottom threshold with 2.8\%, 4.0\%, and 5.1\% relative uncertainty.
The obtained branching fractions for the $\Upsilon$(2S) and $\Upsilon$(3S) are significantly larger than prior measurements and the current world average values, resulting in narrower total decay widths for these resonances.
The new branching fractions, particularly ${\cal B}_{\mu\mu}(2S)$, also affect the measured rates of other transitions leading to the $\Upsilon$ resonances and observed by the subsequent decay $\Upsilon \to \mu^+\mu^-$.

\section{Acknowledgment}
We gratefully acknowledge the effort of the CESR staff 
in providing us with
excellent luminosity and running conditions.
This work was supported by 
the National Science Foundation,
the U.S. Department of Energy,
the Research Corporation,
and the 
Texas Advanced Research Program.


\end{document}